\documentclass[prb,amsmath,amssymb,twocolumn]{revtex4}

\usepackage{graphicx}
\usepackage{color}

\newcommand{\sss}{\scriptscriptstyle}
\newcommand{\STO}{SrTiO$_3$}
\newcommand{\grad}{$^\circ$C\ }
\newcommand{\LCMOF}
{La$_{\sss {2/3}}$Ca$_{\sss {1/3}}$MnO$_{\sss 3}$}
\newcommand{\LCMOA}
{La$_{\sss {1/3}}$Ca$_{\sss {2/3}}$MnO$_{\sss 3}$}
\newcommand{\LCMO}
{La$_{\sss {(1-x)}}$Ca$_{\sss {x}}$MnO$_{\sss 3}$}

\newcommand{\Tc}{T_{\sss{\text{C}}}}
\newcommand{\Tn}{T_{\sss{\text{N}}}}
\newcommand{\He}{H_{\!\sss E}}
\newcommand{\Hmb}{H_{\!\sss E}^{\sss {\text {MB}}}}
\newcommand{\Mf}{M_{\sss{\text {F}}}}
\newcommand{\Sf}{S_{\sss{\text {F}}}}
\newcommand{\tf}{t_{\sss{\text {F}}}}
\newcommand{\Saf}{S_{\sss{\text {AF}}}}
\newcommand{\taf}{t_{\sss{\text {AF}}}}
\newcommand{\Kaf}{K_{\sss{\text {AF}}}}
\newcommand{\muB}{\mu_{\sss{\text B}}}
\newcommand{\Hcl}{H_{ {\rm c}^{}{\sss 1}}}
\newcommand{\Hcr}{H_{ {\rm c}^{}{\sss 2}}}

\begin{document}

\title{Thickness dependence of the exchange bias\\in epitaxial manganite
bilayers}

\author{A.~L.~Kobrinskii}
\affiliation
{
School of Physics and Astronomy, University of Minnesota\\
116 Church St.~SE, Minneapolis, Minnesota 55455, USA
}
\author{Maria Varela}
\affiliation
{
Materials Science and Technology Division, Oak Ridge National Laboratory,
Oak Ridge, Tennessee 37831, USA
}
\author{S.~J.~Pennycook}
\affiliation
{
Materials Science and Technology Division, Oak Ridge National Laboratory,
Oak Ridge, Tennessee 37831, USA
}
\author{A.~M.~Goldman}
\affiliation
{
School of Physics and Astronomy, University of Minnesota\\
116 Church St.~SE, Minneapolis, Minnesota 55455, USA
}

\date{\today}

\begin{abstract}
Exchange bias has been studied in a series of \LCMOF/\LCMOA\ bilayers grown on
(001) \STO\ substrates by ozone-assisted molecular beam epitaxy. The high
crystalline quality of the samples and interfaces has been verified using
high-resolution X-ray diffractometry and Z-contrast scanning transmission
electron microscopy with electron energy loss spectroscopy. The dependence of
exchange bias on the thickness of the antiferromagnetic layer has been
investigated. A critical value for the onset of the hysteresis loop shift has
been determined. An antiferromagnetic anisotropy constant has been obtained by
fitting the results to the generalized Meiklejohn-Bean model.
\end{abstract}



\maketitle

\section{Introduction}

Since the work of Meiklejohn and Bean,\cite{Bean56} structures in which ferromagnetic (F) and antiferromagnetic (AF) materials are juxtaposed, have
been known to exhibit additional unidirectional anisotropy. This anisotropy
was attributed to the exchange interaction between the F and AF spins at the
interface between the materials and manifested itself in hysteresis loop
measurements as a shift of the magnetization curve along the field axis. The
shift is traditionally used to measure the size of the effect and is referred
to as ``exchange bias.'' For the last 50 years, a vast amount of research, both
experimental and theoretical, as well as (more recently) numerical
calculations, has been undertaken to discern the nature of interfacial AF/F
exchange coupling.\cite{RevNandS,RevBandT,RevStamps,RevKiwi,SahaVictora}
A complete theory, which would explain and correctly predict the various
manifestations of exchange anisotropy, is still lacking. However, a number of
models have been proposed and mechanisms have been identified, by which this
anisotropy may occur. The models also predict values of the magnetization
curve shift and enhanced coercivity, their relationship to material
properties, as well as dependencies on the cooling field, temperature, strain,
and thicknesses of AF and F layers. Experiments with AF/F interfaces involving
various measurement techniques have been carried out to quantify the exchange
bias effect in many material systems and structures. For instance, the
experimentally measured dependence of exchange bias on the AF layer thickness
allows one to test theoretical predictions and at the same time is important
from the applications standpoint.

Here, we report measurements of the  dependence of the exchange bias effect in
epitaxial \LCMOF/\LCMOA\ bilayers on the thickness of the AF layer.
Heterostructures of calcium doped lanthanum manganites are particularly
suitable for investigations of interfacial effects such as exchange bias.
Their rich phase diagram allows for the possibility of both ferromagnetic and
antiferromagnetic compounds, which have similar and well matched crystal structures. Growth conditions for these materials are similar, which allows
one to fabricate heterostructures with atomically sharp interfaces.
Exchange-biased manganite multilayers have been grown on LaAlO$_3$ and
\STO\ substrates by laser ablation\cite
{
Panagiotopoulos99jap,Panagiotopoulos99prb,Panagiotopoulos00,Moran}
and by molecular beam epitaxy~(MBE).\cite
{
Eckstein96,ODonnell97,Nikolaev00,Krivorotov01}

\section{Experimental Procedures:               \protect\\
         Fabrication \lowercase{and} Characterization}

The bilayers under study were grown using the method of ozone-assisted MBE in
the block-by-block mode.\cite{Locquet} This approach has been successfully
employed in the fabrication of high quality thin films and heterostructures of
a variety of perovskite oxides.\cite
{
DBCO90,Locquet,Eckstein94,LCMO96,Eckstein96,ODonnell97,Nikolaev00,
Krivorotov01,Dobin03,demo,Eblen-Zayas05,Bhattacharya05}
Our films were grown at 700-750\,\grad on \STO\ (001) substrates, which offer
a good lattice match to \LCMO{$^{}$}.\cite{mismatch} The ozone partial
pressure was $2.0\times10^{-5}\,$Torr while the base pressure was in the
high~$10^{-11}$ to low~10$^{-10}$~Torr range. The details of the growth
apparatus and procedures have been described elsewhere.\cite{HTSC92}
A layer of \LCMOF~(F) about~190\,\AA\ in thickness was deposited first. It was
followed by layers of \LCMOA\ (AF) that were~50 to~250\,\AA\ in thickness
depending upon the sample. The total thickness of each bilayer was kept
below~500\,\AA\ to avoid strain relaxation. The growth rate was
about~8\,\AA/min.~for \LCMOF\ and about~6.5\,\AA/min.~for \LCMOA. The
reflection high energy electron diffraction (RHEED) patterns observed during
growths oscillated with periods corresponding to one monolayer deposition
times. Streaky patterns and the absence of extra features indicated a smooth
growth without the formation of non-stoichiometric phases. Oscillations were
normally stronger during the growth of the ferromagnet.

The crystal quality of the films was also characterized
\begin{figure}
\includegraphics[width=0.46\textwidth]{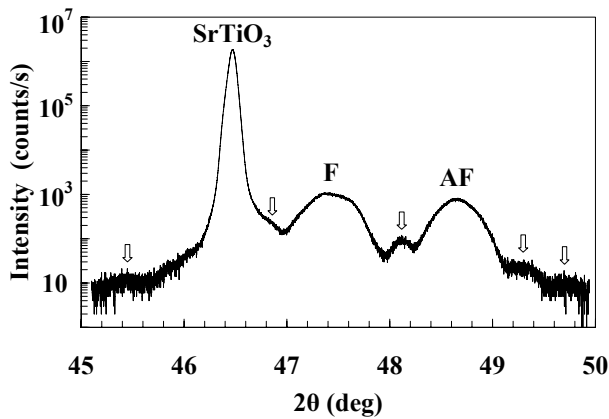}
\caption
{
Symmetric $2\theta$-$\omega$ scan in the vicinity of the (002) peak of
\STO\ substrate. Peaks associated with \LCMOF\ (F) and \LCMOA\ (AF) are shown.
Finite layer size oscillations are marked with arrows.
}
\label{fig:xray}
\end{figure}
using high-resolution x-ray diffractometry~(XRD).
The XRD data obtained from the series of bilayers suggest highly-ordered
epitaxial crystal structures. The two distinct peaks seen in
Fig.~\ref{fig:xray} correspond to two manganite layers (192\,\AA\ F and
250\,\AA\ AF) with slightly different lattice parameters in the direction of
growth. The out-of-plane lattice parameters determined from the positions of
these peaks are~3.84\,\AA\ and~3.76\,\AA\ for \LCMOF\ and \LCMOA,
respectively. This is consistent with the films being under tensile strain set
by the substrate. The full widths at half maxima (FWHM) of the rocking curves
at the (002) Bragg peaks have remarkably low values of
$0.045$\,-\,$0.110^{\circ}$ for all of the films studied. This confirms the
high quality of these samples. In all instances, the FWHM of the F peak was
found to be smaller than that of the AF peak, in agreement with the
\textit{in-situ} RHEED observations. Intensities of the AF peaks vary with
layer thickness, the diffraction signal being at the background level for the
thinnest ($\approx$50\,\AA)\ AF specimens.

In order to check the stoichiometry of the compounds, single (either F or AF)
layer reference samples were grown and analyzed by Rutherford back scattering
and inductively coupled plasma mass-spectroscopy.

\begin{figure}
\centering
\includegraphics[width=0.4\textwidth]{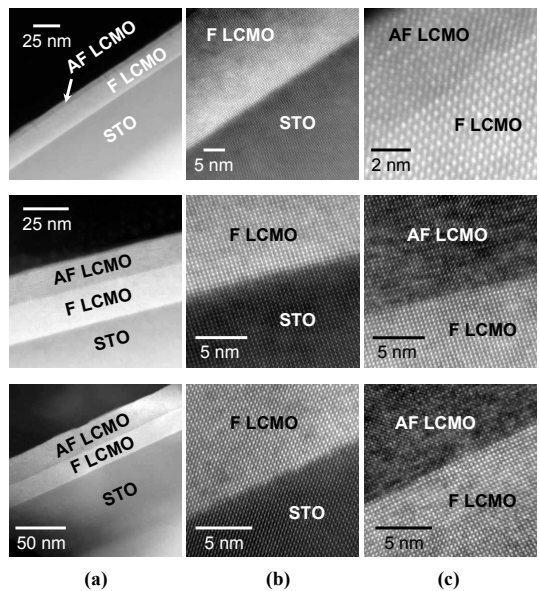}
\caption[Z-contrast STEM images of interfaces]
{
A set of Z-contrast STEM images of 19.2\,nm \LCMOF/$\taf$ \LCMOA~bilayers with
$\taf=5\,$nm (top row), 15\,nm (middle row), and 25\,nm (bottom row).
(a) Low magnification views of both substrate-film and F-AF interfaces.
(b) and (c) Higher magnification images of substrate-film and F-AF interfaces,
respectively.
}
\label{fig:interfaces}
\end{figure}

\begin{figure}[h]
\centering
\includegraphics[width=0.4\textwidth]{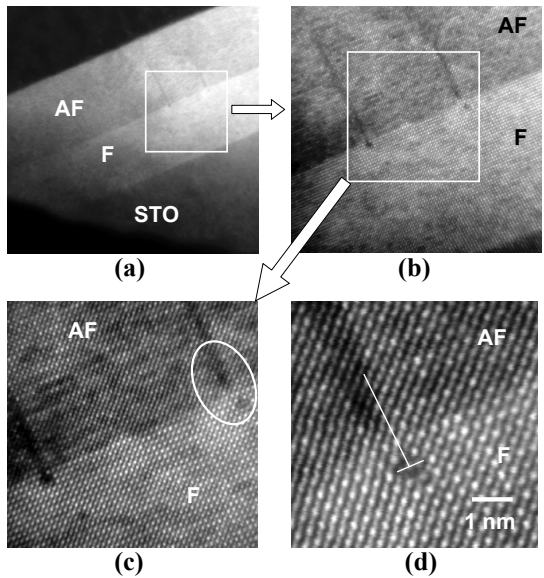}
\caption
{
A set of Z-contrast scanning transmission electron microscopy images showing a
dislocation region in a 19.2\,nm \LCMOF/25\,nm \LCMOA~bilayer. (a) A low
magnification view of a region with two dislocations (boxed).
(b) and (c) Higher magnification views of the dislocations;
(d) High magnification of the dislocation circled in (c)$^{}$. On top of the
dislocation core,
an extra semiplane can be seen that propagates into the AF layer.
}
\label{fig:dislocation}
\end{figure}
Scanning transmission electron microscopy (STEM) studies provided further
insight into the structures under investigation.\cite{EPAPS} The measurements
were carried out in a VG Microscope HB501UX equipped with a Nion aberration
corrector and a Gatan Enfina electron energy loss spectrometer. The microscope
was operated at 100 kV. The Z-contrast images (Fig.~\ref{fig:interfaces})
revealed sharp interfaces between the substrates and the films, as well as
between the F and the AF layers. The layers are flat and
continuous over large lateral distances with occasional monolayer
height steps, and their observed thicknesses are in a good agreement with
the values programmed for growths. The samples are free of major defects.
However, in rare spots (Fig.~\ref{fig:dislocation})$^{}$, dislocations were
observed.

These edge dislocations are possibly due to the mismatch between the film and
the substrate, as well as between the two manganite layers, with an extra
semiplane propagating into the AF layer and serving to relax the strain. We
note that no such defects were observed in the sample with the thinnest AF
layer, which supports the argument, because the strain relaxation of epitaxial
films develops with increasing thickness. The fact that dislocations occur in
bilayers near the interface and propagate into the AF layers is in agreement
with our earlier assertion that the F layers are of higher crystalline quality
than the AF bilayers.

\begin{figure}
\centering
\includegraphics[width=0.47\textwidth]{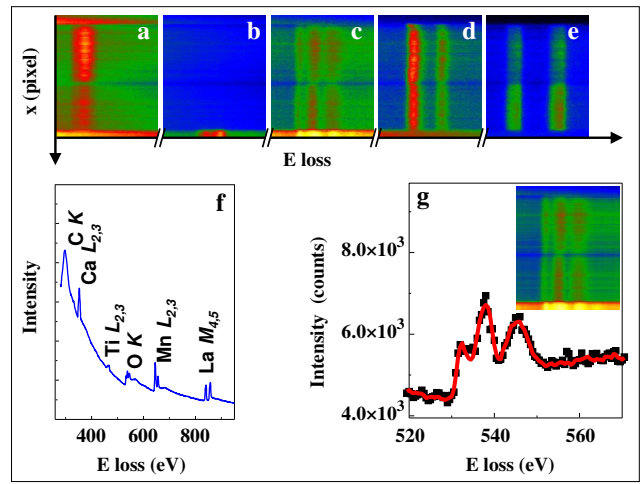}
\caption{
(Color online)
EELS linescan across the AF-F-substrate stack:  Ca\,\textit{L}$_{2,3}$~(a),
Ti\,\textit{L}$_{2,3}$~(b), O\,\textit{K}~(c), Mn\,\textit{L}$_{2,3}$~(d), and
La\,\textit{M}$_{4,5}$~(e) edges. The color scale has been adjusted for
presentation purposes. The acquisition time was one second per spectrum. (f)
Average spectrum for the whole series, showing all the edges acquired
simultaneously, and also the energy scale. (g) Single spectrum showing the
O\,\textit{K} edge, extracted from the F layer, with the black datapoints
corresponding to the raw data and the red line to that treated with the
principal component analysis\cite{Bosman} (PCA). Inset: the same O\,\textit{K}
edge as in (c) upon applying PCA.
}
\label{fig:EELSscan}
\end{figure}

\begin{figure}
\centering
\includegraphics[width=0.47\textwidth]{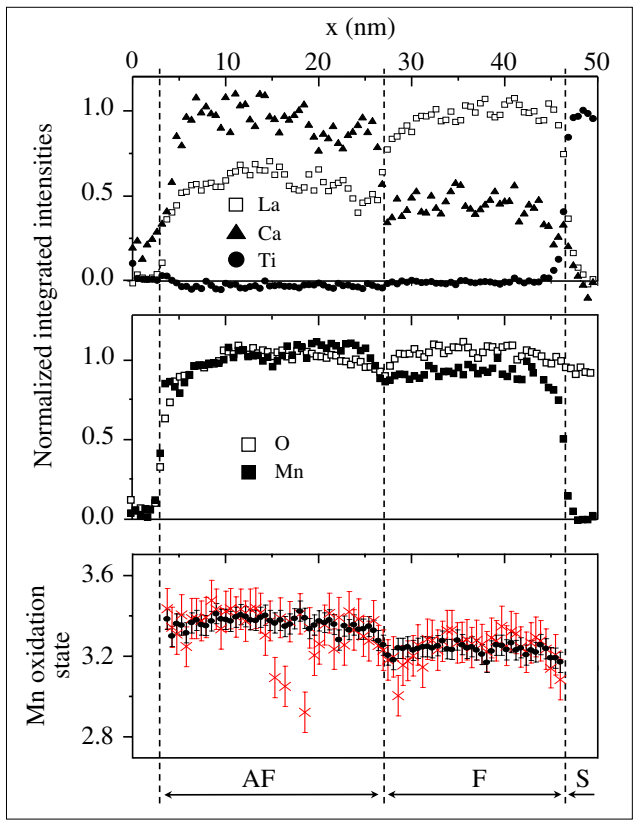}
\caption
{
(Color online)
Elemental profiles in the direction of growth (labeled $x$ here) of the
intensities that correspond to lanthanum, calcium, and titanium (upper graph),
and to oxygen and manganese (middle graph). The bottom graph shows the profile
of the oxidation state of manganese ions across the film: red crosses -- as
measured; black dots -- after applying PCA to
filter out the noise. Regions corresponding to the~\LCMOA\ (antiferromagnetic)
layer, the~\LCMOF\ (ferromagnetic) layer, and the \STO\ substrate are
designated as AF, F, and S, respectively. Elemental profiles were obtained by
subtracting the signal background after a power law fit and integrating the
remaining intensity under the edge of interest (Ca\,\textit{L}, O\,\textit{K},
Mn\,\textit{L}, and La\,\textit{M}). This integrated intensity was then
normalized. The Mn oxidation state was obtained by measuring the pre-peak
intensity of the O\,\textit{K} edge near~530\,eV. The measurements were
carried out at room temperature. The data were averaged over a length of~2\,nm
in the direction parallel to the interface.
}
\label{fig:EELSdata}
\end{figure}

Along with STEM imaging, electron energy loss spectroscopy (EELS)
measurements were carried out, which allowed us to look closely at the
chemical composition of the samples.\cite{EPAPS}
Figure~\ref{fig:EELSscan} shows the electron energy loss spectra acquired
while scanning the electron beam across the structure.
The Ca\,\textit{L}$_{2,3}$, Ti\,\textit{L}$_{2,3}$, O\,\textit{K},
Mn\,\textit{L}$_{2,3}$, and La\,\textit{M}$_{4,5}$ edges are shown.
Panel (f) shows the average spectrum for a full line scan. Principal
component analysis\cite{Bosman} (PCA) was applied to remove random noise. Panel
(g) expands the O\,\textit{K} region of an energy loss spectrum
from the F layer. The inset shows the O\,\textit{K} edge portion of the line
scan after PCA. The agreement between the raw data and the PCA treated data is
excellent. Figure~\ref{fig:EELSdata} shows the elemental EELS integrated
intensity profiles across a bilayer. The profiles are consistent with
atomically sharp interfaces, of course considering that the specimens
are thick in the electron beam direction and some electron beam
broadening is to be expected. The upper panel in Fig.~\ref{fig:EELSdata} shows
that the compositional change occurs within a region as narrow as two unit
cells (a nm or less) for the AF-F interface. The substrate-film
interface appears to be broader, with a width under 2\,nm (4-5 unit
cells). This effect is most likely due to an enhanced beam broadening
due to the fact that the STEM specimen was considerably thicker in
this area. Nevertheless, this interface has no bearing on the main
subject of this paper: the effect of the AF layer thickness on the
exchange bias properties.

The properties of manganites are very sensitive to even small variations in
both the doping level and the degree of oxygenation. As the structure of a
sample is scanned in the direction of growth in the STEM-EELS study, calcium
and lanthanum signals are expected to change at the bilayer interface by a
factor of~2$^{}$, while those of oxygen and manganese should remain at a
constant level throughout the film. Our data are in a reasonably good agreement with these predictions.

The Mn oxidation state can be quantified from the analysis of the fine
structure of the O K edge around~530\,eV. It can be obtained from the
measurement of the difference in energy between the pre-peak feature
and the adjacent main peak of the O\,K edge. This difference, $\Delta E$,
increases linearly with Mn oxidation state in~\LCMO$^{}$. A linear
fit can be obtained from this \LCMO\ series and it can be used as a
calibration for obtaining Mn valences from $\Delta E$ measurements.\cite{Maria}
Principal component analysis was used to remove random
noise. Both the result of the analysis of raw EELS data
and the PCA-treated data are plotted in Fig.~\ref{fig:EELSdata}$^{}$.

Based on the information available from the EELS experiments, we conclude that
manganese valence is reduced as compared to its nominal stoichiometric values
of~$+3.3$ and~$+3.6$ for the F and AF compounds, respectively, which we
attribute to possible oxygen vacancies and the lanthanum content being
slightly above the target level within the AF layer (0.5 on the upper graph of
Fig.~\ref{fig:EELSdata}). We also note that within several unit cells of the
surface of \STO~substrate, manganese oxidation state drops down to about
$+3.1$\,. This suggests the presence of a thin ($\lesssim2\,$nm) magnetically
dead layer. Interestingly, this was not the case in ultrathin films of ferromagnetic La$_{\sss {0.8}}$Ca$_{\sss {0.2}}$MnO$_{\sss 3}$ grown by
codeposition.\cite{Eblen-Zayas05,Bhattacharya05}

The results of the structural characterization demonstrate that the bilayers
under study are of a very high crystalline quality, with precisely calibrated
thicknesses of both layers. Grown without a seed or a cap layer, they preserve
the stoichiometry and oxygen content, and as well are free of major defects
and additional phases. The films are chemically clean and possess physically
and chemically sharp interfaces, which makes them a good system for a study of
the exchange bias effect.

\section{Experimental procedures:               \protect\\
         Physical Measurements                                 }

\begin{figure}
\centering
\includegraphics[width=0.47\textwidth]{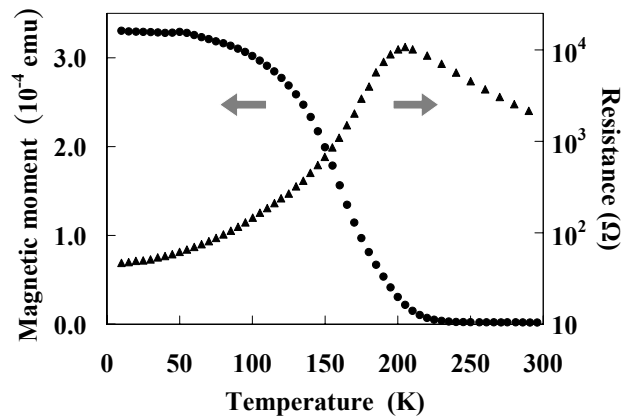}
\caption
{
Temperature dependence of magnetization (circles) and resistance (triangles)
of a F\,192\,\AA/AF\,150\AA\ bilayer. The same in-plane magnetic field
of~100\,Oe was applied during both cooling and measurement processes. This
result is typical of many samples. We note an anomaly at about 55\,K in the
magnetization data which, remarkably, was observed in all samples at the same
temperature. While we do not have an explanation of this effect, we would like
to point out that the dielectric constant of the \STO\ substrate is known to
increase rapidly at low temperatures taking off at about the same temperature
as the anomaly. To what extent this may affect the properties of the
ferromagnet has not been studied in the present work.
}
\label{fig:MandRvsT}
\end{figure}

The resistances of the films were measured using a Quantum Design Physical
Properties Measurement System. Samples were cooled from room temperature down
to~10K in an in-plane field of 100\,Oe and the resistance was measured on
warming in a four-point van der Pauw geometry. The right-hand graph in Fig.~\ref{fig:MandRvsT} shows the temperature dependence of the resistance of
a bilayer sample. The conductivity of the bilayer is determined by that of the
ferromagnet, which undergoes a transition from a metallic to an insulating
state at a temperature slightly below 200K.

Magnetic properties of the bilayers were studied using a Quantum Design
Magnetic Properties Measurement System. Magnetization was measured as a
function of temperature on warming, after cooling a sample from room
temperature to~10\,K in a magnetic field. The field was applied along the~[100]
direction in the plane of the sample. Figure~\ref{fig:MandRvsT} shows
the ferromagnetic-to-paramagnetic transition in a
F\,192\,\AA/AF\,150\,\AA\ bilayer. The onset of ferromagnetism is coupled with
the metal-to-insulator transition as expected from colossal
magnetoresistive~\LCMOF. For all of the samples studied, values of the Curie
temperature,~$\Tc$, determined by the inflection points of temperature
dependence graphs, were found to be about~175\,K, which is lower than that of
bulk material.\cite{WollanKoehler} Below 200\,K, there appears a well-defined
transition followed by a low-temperature plateau region. We note that this is
a behavior different from that seen in magneto-thermal curves obtained for
field-cooled Ca-doped lanthanum manganite heterostructures grown by laser
ablation on LaAlO$_3$
substrates.\cite{Panagiotopoulos99jap,Panagiotopoulos99prb,Panagiotopoulos00}
The latter exhibit monotonic behavior without saturation of the magnetic
moment at low temperatures. The behavior shown in Fig.~\ref{fig:MandRvsT}
is similar to that reported in the work by Mor\'an {\it et al}.~for films and
superlattices grown on single crystal
(001)-(LaAlO$_{3}$)$_{0.3}$(Sr$_{2}$AlTaO${_6}$)$_{0.7}$ substrates using
pulsed laser deposition.\cite{Moran} Quantitatively, upon cooling in an applied
field of 100\,Oe, we observe a low-temperature magnetization of about~500\,emu/cm$^3$, which is significantly greater than the values
reported\cite{Panagiotopoulos99prb,Moran} for laser ablated films cooled
in~10\,kOe. 
\begin{figure}
\centering
\includegraphics[width=0.47\textwidth]{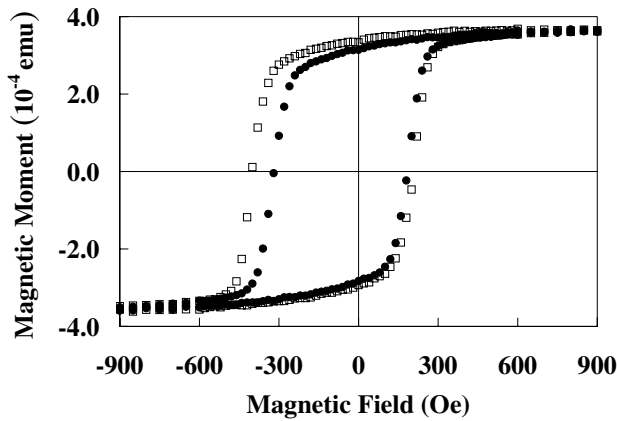}
\caption{
The first (open squares) and the second (solid circles) magnetization
hysteresis curves for an exchange-biased F\,192\,\AA/AF\,150\,\AA\ bilayer
taken at~10\,K. The data show a training effect similar to that observed in
other systems~\cite{Hoffmann} and reproduced in numerical
calculations.\cite{SahaVictora}$^{}$
}
\label{fig:Hyst}
\end{figure}

The N{\'e}$^{}$el temperature $\Tn$ of bulk \LCMOA\ found by Herrero
\textit{et al.}\cite{Herrero} is~170\,K. This value can be taken as the upper
limit for $\Tn$ in thin films. In order to set the exchange bias, bilayers
were cooled from room temperature through $\Tc$ and, subsequently, $\Tn$, in
a~500\,Oe in-plane magnetic field. A typical hysteresis loop is presented in
Fig.~\ref{fig:Hyst}. It was determined from magnetization curves taken at
higher temperatures that the hysteresis loop shift vanishes at about 90\,K.
This is in a good agreement with the previously reported results
for~\LCMOF/\LCMOA\ heterostructures.\cite{Krivorotov01}

The low temperature magnetic moment (Fig.~\ref{fig:MandRvsT}) is close to
that measured in a high magnetic field (Fig.~\ref{fig:Hyst}),
which indicates that the magnetic structure of the F layer is close to single
domain. For La$_{1-x}$Ca$_{x}$MnO$_3$ compounds with Ca doping
levels~$x\approx0.3\,$, the following expression\cite{Salamon} can be used to
estimate the average low temperature magnetic moment $m_{\sss{\text S}}$ per
chemical formula unit (i.~e.~per Mn) as the total of the contributions expected
from $x$ Mn$^{3+}$ and~$(1-x)$~Mn$^{4+}$ ions.
\begin{equation}
\label{ms}
m_{\sss{\text S}} = 4x\muB + 3(1-x)\muB=(3+x)\muB\,.
\end{equation}
Given that the volume of our ferromagnetic layer is approximately~$7.5\times10^{-7\,}$cm$^3$, we obtain a low temperature
magnetization of about $4.0\times10^{-4\,}$emu. The saturation magnetic moment
measured to be~$3.7\,$-$\,3.8\times10^{-4\,}$emu in all studied bilayers is in
a good agreement with that calculated.

Consecutive field sweeps resulted in different coercivities for the first and
the second loops only, with no noticeable changes occurring in further cycles.
Training effects have been observed in a number of exchange bias systems
utilizing different AF materials.\cite{training} We note (Fig.~\ref{fig:Hyst})
that similar to other systems,\cite{Hoffmann} the effect is stronger on the
left coercive field. However, our data (Fig.~\ref{fig:Hyst}) are different in
that the shape of hysteresis loops is much the same for the first and the
subsequent field sweeps. The data on the thickness dependence of exchange bias
discussed next were obtained by using the values of the ``trained'' coercive
fields\cite{crystallinity} $\Hcl$ and $\Hcr\,$ to calculate the value of
exchange bias as~$H_{{\rm ex}}=\tfrac{\Hcl+\Hcr}{2}\,$.

\section{Results and Analysis}

\begin{figure}
\centering
\includegraphics[width=0.47\textwidth]{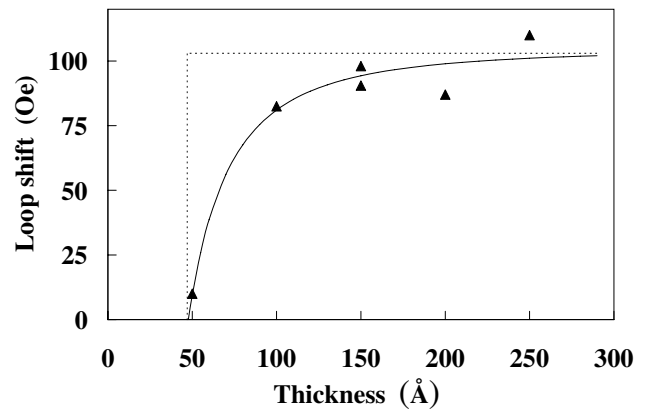}
\caption
{
Exchange bias as a function of AF layer thickness for a series of
\LCMOF\,192\,\AA/\LCMOA\ bilayers at~10\,K. Solid line shows the generalized
$\He(\taf)$ of Eq.~\eqref{eq:Fit}. The limit of $\Kaf\!\to\!\infty$ (dashed
line) corresponds to the original MB model.}
\label{fig:ThicknessDependence}
\end{figure}

The results of our study of a series of bilayer films are presented by the
plot in Fig.~\ref{fig:ThicknessDependence}. The dependence shown is
characterized by the saturation of exchange bias~$\He$ with the
antiferromagnetic layer thickness~$\taf\,$ and the existence of a critical
value $\taf^{\text{cr}}\,$, below which no hysteresis loop shift occurs.
In the first theoretical model proposed by Meiklejohn and Bean (MB
model),\cite{Bean57,Meiklejohn62} the magnitude of the exchange bias~$\Hmb$
does not depend on $\taf$ and is estimated from
\begin{equation}
\label{eq:MB}
{\Hmb} = \frac{ J^{} \Sf \, \Saf }
              { \Mf^{} \, \tf^{} }
\,,
\end{equation}
where $S_{\sss{\text {AF/F}}}$ is spin of AF/F ions at the interface, $J$ is
their exchange coupling constant, and $\Mf$ and~$\tf$ are the ferromagnetic
layer magnetization and thickness, respectively.
Although there is no AF thickness dependence in~Eq.\,\eqref{eq:MB}, the
exchange bias is expected to vanish when the anisotropy energy in the AF layer
is not large enough to keep AF spins from reversal caused by their exchange
interaction with the interfacial F spins, as the latter switch in an applied
magnetic field, i.~e.
\begin{equation}
\label{eq:CrTh}
\Kaf^{}\,\taf^{\text{cr}}=
J^{} \Sf^{} \Saf^{}
\,,
\end{equation} 
where $\Kaf$ is the antiferromagnetic anisotropy constant. This relationship
has been used before\cite{Mauri} to estimate~$\Kaf$\,.

Binek \textit{et al}.\cite{Binek}~generalized the simple model of Meiklejonh
and Bean and derived an equation, which takes into account the effect of finite
magneto-crystalline anisotropy of the AF layer. According to their result, in
the case of finite but strong $\Kaf$, the exchange bias depends on the AF layer
thickness as
\begin{equation}
\label{eq:Fit}
    \He^{}
=   \He^{\infty}
    \left( 1- \frac{ {(\Delta E)}^{2} }{ 8^{} \Kaf^2 \taf^2 }
    \right)
,
\end{equation}
where $\Delta E \equiv J^{} \Sf \Saf$ is an interface energy per unit area and
the saturation value of the exchange bias at large~$\taf$ is
$\He^{\infty}\equiv\Delta E/(\Mf^{} \tf^{})=\Hmb$.
This model does not explain the nature of the AF/F interfacial exchange
coupling -- the product~$J^{} \Sf \Saf$ enters the quantitative analysis as a
phenomenological interfacial energy. The energy~$\Delta E$ is a convenient
measure of the effect for it does not depend on the nature of the~F~layer
material, nor on its thickness $\tf\,$, thus allowing one to compare different
exchange-biased systems. From the data in Fig.~\ref{fig:ThicknessDependence}
we take $\He^{\infty}\approx100\,$Oe and
estimate~\mbox{$\Delta E = \He^{\infty}\Mf^{} \tf^{}\approx0.1$\,erg/cm$^2$},
which is within the range of values reported for \LCMOA\ and other
antiferromagnetic materials.\cite{RevNandS,Panagiotopoulos00}

Equation~\eqref{eq:Fit} can be used to fit experimental data for the AF
thickness dependence. Lund \textit{et al}.~recently
demonstrated\cite{Leighton02} that this fitting procedure does not meet with
equal success in systems with different anisotropies of the AF layer: it works
reasonably well for MnF$_2$/Fe and fails in the stronger anisotropy FeF$_2$/Fe
system. The dependence~$\He(\taf)$ of Eq.~\eqref{eq:Fit} is plotted in
Fig.~\ref{fig:ThicknessDependence} to show that the generalized Meiklejohn-Bean
model satisfactory explains the rise and saturation features of the AF
thickness dependence. Quantitatively, as compared to the
condition~\eqref{eq:CrTh} for the critical thickness, an additional numerical
factor is introduced. Setting $\He=0\,$ in~Eq.~\eqref{eq:Fit}, we
get\;$2\sqrt{2} \, \Kaf^{}\, \taf^{\text{cr}} = \Delta E\,$. From the data, the
critical thickness is between 50 and 100\,\AA, based on which the AF anisotropy
is~$4.0 \div 7.0\times10^{4\,}$erg/cm$^3$.

As far as the assumptions made in the generalized MB approach to obtain the
simple $\taf$ dependence of the form shown in Eq.~\eqref{eq:Fit}, our epitaxial
bilayers and the measurement set-up are similar to the model system. However,
there are several differences. While uniaxial anisotropy and coherent rotation
are assumed in the MB model, calcium doped lanthanum manganite films on
\STO\ (001) substrates possess a biaxial in-plane anisotropy. Magnetization
reversal in such films was shown\cite{ODonnell97} to occur through domain wall
motion. As compared to magnetization curves observed in our experiment (see
Fig.~\ref{fig:Hyst}), magnetization reversal by coherent rotation in a system
with a uniaxial anisotropy would likely result in a more square shape of the
hysteresis loop. This, as well as some training, affects the values of the loop
shift as they are calculated using the coercive fields taken from the measured
magnetization curves. One can see from the data of Fig.~\ref{fig:Hyst}\,,
however, that the reversal still occurs within a narrow range of field and thus
the F layer state may be considered to be close to a single domain through most
of the sweep cycle.

\section{Conclusion}

Using the method of block-by-block ozone-assisted molecular beam epitaxy, we
have been able to fabricate high quality epitaxial F/AF bilayers of lanthanum
manganites by modulating the level of calcium substitution. We have obtained
the dependence of the exchange bias on the AF layer thickness and determined
the critical thickness $\taf\,$. The vanishing of the effect below a critical
AF layer thickness and its saturation at large AF layer thicknesses is
described relatively accurately within the framework of the generalized model
of Meiklejohn and Bean. Using the model, we estimate the low temperature
interfacial energy of exchange interaction to
be~\mbox{$\approx0.1\,$erg/cm$^2$} and the AF anisotropy to
be~\mbox{$\Kaf\approx7.0\times10^{4\,}$erg/cm$^{3}$}. To the best of our
knowledge, the value of the anisotropy energy of antiferromagnetic \LCMOA\ has
not been previously reported. In addition to the classical methods, such as
torque magnetometry\cite{Uchida} or neutron diffraction,\cite{Hutchings} a
torque magnetometry technique utilizing anisotropic magnetoresistance was
developed and successfully employed\cite{Dahlberg88,Dahlberg96,Krivorotov02}
more recently for studies of exchange anisotropy in epitaxial exchange-biased
systems. Further experiments are in order to study the exchange anisotropy of
the antiferromagnetic manganite \LCMOA~in greater detail.

\section*{\uppercase{Acknowledgments}}

The authors would like to thank Konstantin Nikolaev, Dan Dahlberg, Alexander
Dobin, Ilya Krivorotov, Chris Leighton, and Jyotirmoy Saha for useful
conversations. They would also like to thank Masaya Nishioka for technical
assistance. The authors are grateful to J.~T.~Luck for helping with STEM
specimen preparation, to M.~Oxley for performing dynamical simulations of
electron scattering, and to M.~Watanabe for providing a plug-in to carry out
PCA in Digital Micrograph. This work was supported by the National Science
Foundation through the University of Minnesota Materials Research Science and
Engineering Center under Grant NSF/DMR-0212032. Research at ORNL was sponsored
by the Division of Materials Sciences and Engineering of the US Department of Energy.


\begin{thebibliography}{99}
\bibitem{demo}
Films of La$_{0.8}$Ca$_{0.2}$MnO$_3$ as thin as~82\,\AA, deposited directly on
\STO\ substrates, were found to undergo a transition to a metallic
state.\cite{Eblen-Zayas05} This confirms the capability of our deposition
system to produce very high quality films of doped colossal magnetoresistance
manganites.
\bibitem{mismatch}
The pseudocubic lattice constants are~3.86\,\AA\ and~3.78\,\AA\ for
bulk\cite{WollanKoehler} \LCMOF\ and \LCMOA, respectively, which results in a
maximum mismatch of 2\% with \STO\ substrates.
\end{thebibliography}
\end{document}